\documentclass[12pt]{article}
\usepackage{amsmath}
\usepackage{amssymb}
\usepackage{graphicx}
\usepackage{fancybox}
\usepackage{lineno}
\usepackage{color}
\usepackage{natbib}

\usepackage{setspace}

\usepackage[labelfont=bf,labelsep=period,justification=raggedright]{caption}


\date{}

\newcommand{\be}{\begin{eqnarray}}
\newcommand{\ee}{\end{eqnarray}}

\title{Risk aversion as an evolutionary adaptation}
\author{Arend Hintze$^{1,3,\ast}$, 
Randal S.~Olson$^{2,3}$, 
Christoph Adami$^{1,3}$,
Ralph Hertwig$^{4}$
\\
$^{1}$Department of Microbiology and Molecular Genetics
\\
$^{2}$Department of Computer Science and Engineering\\
$^3$BEACON Center for the Study of Evolution in Action\\
Michigan State University, East Lansing, MI, USA\\
\\
$^4$Center for Adaptive Rationality\\
Max Planck Institute for Human Development, Berlin, Germany
\\
$\ast$ E-mail: hintze@msu.edu}

\begin{document}
\maketitle

\section*{Abstract}
Risk aversion is a common behavior universal to humans and animals alike. Economists have traditionally defined risk preferences by the curvature of the utility function. Psychologists and behavioral economists also make use of concepts such as loss aversion and probability weighting to model risk aversion. Neurophysiological evidence suggests that loss aversion has its origins in relatively ancient neural circuitries (e.g., ventral striatum). Could there thus be an evolutionary origin to risk avoidance? We study this question by evolving strategies that adapt to play the {\em equivalent mean payoff gamble}. We hypothesize that risk aversion in the equivalent mean payoff gamble is beneficial as an adaptation to living in small groups, and find that a preference for risk averse strategies only evolves in small populations of less than 1,000 individuals, while agents exhibit no such strategy preference in larger populations. Further, we discover that risk aversion can also evolve in larger populations, but only when the population is segmented into small groups of around 150 individuals. Finally, we observe that risk aversion only evolves when the gamble is a rare event that has a large impact on the individual's fitness. These findings align with earlier reports that humans lived in small groups for a large portion of their evolutionary history. As such, we suggest that rare, high-risk, high-payoff events such as mating and mate competition could have driven the evolution of risk averse behavior in humans living in small groups. 
   

\section*{Introduction}
When people are faced with dicey decisions, a well-documented trend holds~\citep{Bernoulli1738,Pratt1964,Arrow1965}: If the stakes are sufficiently high, people are risk averse. Risk averseness is usually described as a resistance to accept a deal with risky payoff as opposed to one that is less risky or even safe, even when the expected value of the safer bargain is lower. This tendency can be explained in two ways. First, the classical economists' account of risk aversion is in terms of the shape of the utility function in expected utility theory. Specifically, the curvature of the utility function is interpreted to measure the agent's risk attitude, and the more concave the utility function, the more risk averse the agent will be. A concave utility function corresponds to the notion of diminishing marginal utility of wealth according to which ``the additional benefit which a person derives from a given increase of his stock of a thing, diminishes with every increase in the stock that he already has"~\cite[p. 79]{Marshall1920}. Second, cumulative prospect theory, perhaps the most influential descriptive account of decision making under risk choice among psychologists and behavioral economists, models risk aversion in terms of three different but related concepts: diminishing marginal utility, loss aversion (i.e., the pain of losses is felt stronger than the joy of equivalent gains), and probability weighting (i.e., the elevation of the weighting function and thus the degree of over/underweighting of small probabilities of gains and losses, respectively)~\citep{Kahneman1979,TverskyKahneman1992}.

Risk aversion is also observed outside of economic decisions. For example, foraging animals actively avoid foraging in an area if they cannot reliably find food there~\citep{Smallwood1996,KachelnikBateson1996}. This work suggests that the risk attitude of both humans and animals may be shaped by a common fundamental principle~\citep{MarshKacelnik2002}. Moreover, there is considerable evidence from cognitive neuroscience that loss aversion has a neural basis~\citep{Trepeletal2005}. Neurophysiological measurements suggests that different regions of the brain process value and risk assessment~\citep{Christopoulosetal2009,Symmondsetal2010}. This work also implies that the neural circuitry that encodes risk aversion (or its building blocks such as loss aversion) is phylogenetically ancient. However, the origin of this neurally encoded risk aversion is rarely discussed (but see~\cite{Okasha2007,Schulz2008,Stern2010}), even though there is strong evidence that risk-taking behavior has significant genetic components~\citep{Cesarinietal2009,Bell2009}. Here, we suggest that selective pressures acting during the evolution of human populations in the past can explain risk averse behavior.

To test whether people are risk averse or risk prone, subjects are usually presented with a monetary gamble where there is a safe choice and a risky choice. For example, the safe choice rewards the subject with a fixed payoff $P$ with 100\% certainty, whereas the risky choice rewards the subject with a higher payoff $Q$ ($Q > P$) only half of the time. In this {\em equivalent mean payoff gamble}~\citep{Silberberg1988}, the rewards are designed such that the mean payoff of the safe and the risky choice is the same. In other words, as long as the same choice is repeated during the lifetime or over evolutionary time, both risk prone and risk averse choices have the same payoff in the long run. Thus, no choice should be preferred. 

In contrast to the {\em equivalent mean payoff gamble} studied here, it has been shown~\citep{Clark1993,Yoshimura1991,Robson1996} that if the risky choice comes with a higher payoff than the certain one, evolution will favor risk prone behavior. Similarly, the certain choice is favored if it has a higher payoff than the risky one. While geometric mean fitness maximization or bet hedging can influence the payoffs received~\citep{Donaldson2008}, these strategies do not affect the payoff in the {\em equivalent mean payoff gamble}. Geometric mean fitness maximization and bet hedging requires multiple games, which is not allowed in the gamble studied here. Utility theory also does not explain risk aversion because it makes the same prediction for the certain and risky choice because their payoffs are identical. While prospect theory correctly predicts the human choice, it does not provide a reason for why these choices are beneficial or adaptive, which is the focus of this study.

Of course, the gamble described above is a crude simplification of human choice under risk, which can be shaped by many other factors. External factors such as framing~\citep{Tversky1981}, how the odds are presented~\citep{Hoffrage2000}, or if the decision has to be made from experience or from description~\citep{Hertwig2009} play a role in human decision making under risk. The relative value of the payoff to the subject as well as whether the gamble is real or hypothetical can have an effect on the subject's preference~\citep{Holt2002}. Internal factors such as age~\citep{Harbaugh2002,Levin2003}, cognitive ability~\citep{Boyle2011}, and habits or personal circumstance~\citep{Campbell1999,Constantinides1990} influence the subject's decision, as well as how the subject weighs the potential value of losses and gains~\citep{Kahneman1979}. Without downplaying the importance of these factors, the question remains: Where did this risk averse behavior originate from?

Intuitively, one would argue that risk-seeking behavior is not favored by evolution. In fact, it has been proposed that animals actively avoid risk due to the increased mortality risky decisions often entail~\citep{Stern2010}. When foraging, animals only take risks when the risk of the decision is outweighed by other factors~\citep{Bateson2002,Bednekoff1996,Houston1991,Poethke2008}. Additionally, foraging animals avoid risk when resources are plentiful, but adopt riskier strategies when resources are scarce~\citep{Stephen1986}. Further, organisms ranging from bacteria~\citep{Beaumont2009} to birds~\citep{Bulmer1984} are suggested to mitigate risk in their reproductive success via {\it bet hedging}~\citep{Gillespie1974,Slatkin1974,Philippi1989,Lehmann2007}. However, in these natural situations risky behavior often does not compensate for the potential cost of taking the risk. Thus, while these circumstances could explain the evolutionary utility of risk averse behavior in many natural settings, it does not explain why humans would be prone to risk aversion in equivalent mean payoff gambles.

Previous studies have reported that in small populations of evolving organisms, the fitness of riskier behaviors is significantly affected by the variance in the payoff of the behavior~\citep{Gillespie1974,Gillespie1977}. This observation suggests that strategies that minimize the variance in the payoff of a gamble should have a selective advantage only in small populations. Consequently, evolution in small populations could potentially explain the origin of risk aversion by humans in equivalent mean payoff gambles. Throughout evolutionary history, humans have experienced at least two population bottlenecks that reduced the human population to as few as 1,000 individuals~\citep{Cann1987,Vigilant1991}. However, a population size of 1,000 individuals is unlikely to be small enough to evolve risk averse behavior as a dominant strategy in the population~\citep{Gillespie1974,Gillespie1977}. Instead, a more likely explanation is that humans have lived in groups of about 150 individuals throughout their evolutionary history~\citep{Aiello1993,Dunbar1993}, which plausibly could have been a small enough effective population size for risk aversion to have evolved.

To test whether evolution can explain the emergence of risk averse strategies, we generalize the equivalent mean payoff gamble (with a safe and a risky choice) so that there are an infinite number of possible choices, parameterized by the probability $\chi$ to obtain the high payoff. We choose this payoff to be $1/\chi$, so that the mean payoff of {\em any} choice will be 1. We will call any of the possible gambles a {\em strategy}, and denote each strategy by the probability $\chi$. The choice $\chi=1$ then implies that the agent chooses the safest gamble. In this game, there is no limit on how risky the gamble is, except we do not allow strategies with $\chi=0$, as they are not normalizable.

In order to study the evolution of the strategy, we simulate a population of agents whose choice of strategy is determined genetically, and inherited by the agent's offspring. The payoff that the agent receives is taken as the agent's fitness.
A small probability of mutation introduces variation, so that alternative strategies from the ancestral one can be explored. Each agent makes only one decision during its lifetime that determines its fitness, which means that the agents are potentially making a life (positive payoff $1/\chi$) or death (zero payoff) decision. Such a life or death decision is akin to a rare lifetime event that has a large impact on an individual's fitness, such as mating and mate competition~\citep{Buss1993}. We use this agent-based simulation to explore how small a single population has to be in order to have a significant impact on the evolution of risk aversion. Additionally, we implement an island-based model to test whether larger populations that were segmented into small groups (with the possibility of migration between groups) could still select for the evolution of risk aversion. Although this model cannot take into account the complexity of human evolution nor the exact circumstances thereof, it can address the plausibility of risk aversion as an evolutionary adaptation to equivalent mean payoff gambles due to small population sizes.

\section*{Results}

\subsection*{Evolution in a single population}

Each agent in a population is represented by a single probability $\chi$ (the agent's inherited gambling strategy), where $\chi$ determines the fitness of the agent. Every agent only plays the gamble once in their lifetime, so their fitness is determined by polling a random variable $X$ 
\be
X=\left\{    \begin{matrix} 
      \frac{1}{\chi} & p= \chi \\
      0 & p=1-\chi\\
   \end{matrix}\right. \label{var1}\;.
\ee

exactly once, where $p$ is the probability to receive the corresponding payoff and $\chi$ is the agent's strategy. An agent equipped with a strategy $\chi>0.5$ is considered risk averse, whereas an agent with a strategy $\chi<0.5$ is considered risk-prone. All else being equal, we expect that evolution should not prefer any strategy over another, because the mean payoff of a species (individuals with the same $\chi$) should be the same regardless of $\chi$.

\begin{figure}
\centering
\includegraphics[width=6.5in]{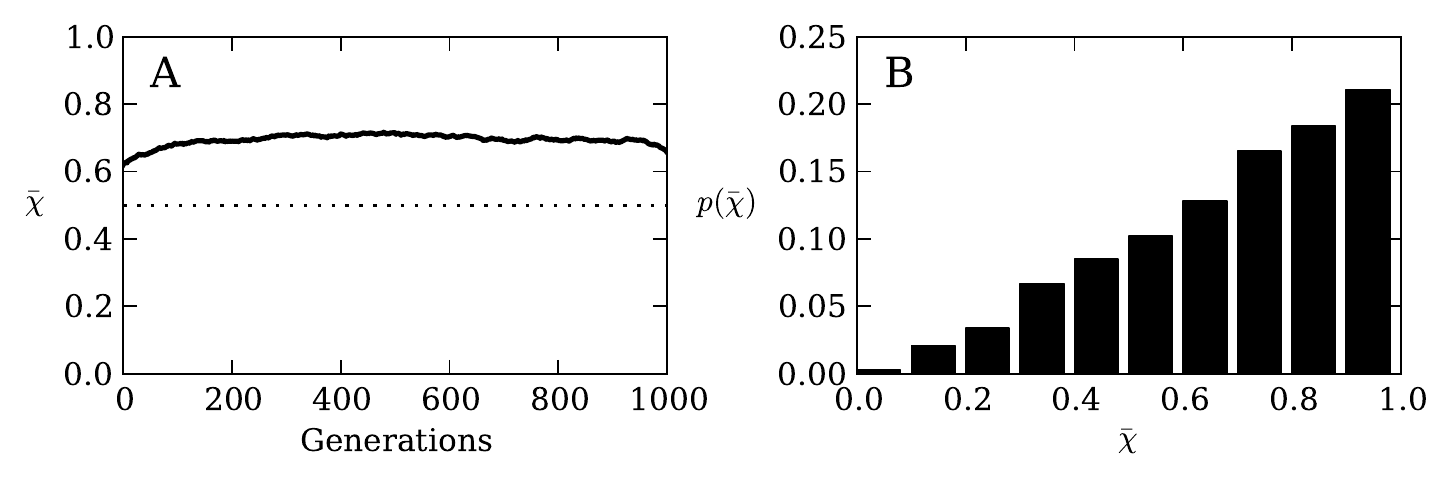}
\caption{Strategy evolution with a fixed population size of 100 individuals and a mutation rate of 1\%. A) Mean strategy $\bar\chi$ on the line of descent at generation 950 over 1,000 replicate runs. Measurements were taken after selection happened, hence the value for generation 1 is not $0.5$. The agents on the line of decent show a preference for risk averse strategy. The dotted line indicates the expected value $0.5$ for unbiased evolution, i.e., no strategy preference. B) The probability distribution of $\chi$  at generation 950 of the dominant strategy across 1,000 replicate runs. This is identical to the distribution of strategies within the population at generation 950 (Figure S2), showing that there is no considerable difference between line of descent and population averages.  The agents evolve a significant preference for risk averse strategies by generation 950 (Wilcoxon rank sum test P $= 7.7954e^{-22}$ between this distribution and a uniform random distribution).}
\label{fig:figSingleLODpop1000beta1}
\end{figure}

If population size does not have a significant effect on the evolution of risk aversion, we would expect the strategy preference of any individual to drift neutrally, so that at the end of the evolutionary run (generation 950) the expected mean population strategy is $\bar\chi=0.5$ (the mean of a uniformly distributed random variable constrained between zero and one).  Instead, we observe in Figure~\ref{fig:figSingleLODpop1000beta1}A that the mean $\chi$ converges to $0.6941\pm0.0139$ (mean $\pm$ two standard errors). 

Similarly, we would expect that if the strategy drifts neutrally, we should observe $\chi$ to be distributed in a uniform manner at the end of the evolutionary runs. Instead, for a population size of $N=100$, we observe in Figure~\ref{fig:figSingleLODpop1000beta1}B a distribution that departs significantly from uniformity (Wilcoxon rank sum test P $= 7.7954e^{-22}$ between this distribution and a uniform random distribution). This result suggests that population size plays a critical role in shaping what strategies evolve in the agent population. 

\begin{figure}
\centering
\includegraphics[width=0.45\textwidth]{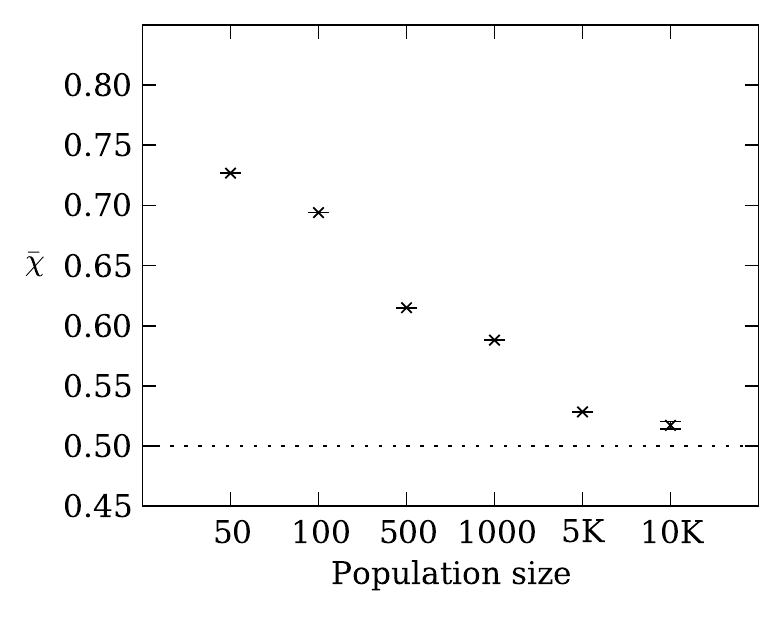}
\caption{Mean strategy $\bar \chi$ at the end of 1,000 evolutionary runs as a function of population size. Agents in smaller populations (e.g., 50 and 100) demonstrate a clear preference for risk averse strategies. In contrast, agents in larger populations (e.g., size 5,000 and 10,000) display only weak risk aversion or no preference. Error bars are two standard errors over 1,000 replicates. The dotted line indicates the expected mean value for unbiased evolution, i.e., no strategy preference. \label{fig:fig2}}
\end{figure}

To further explore the effect of population size on evolved strategy preference, we ran the evolutionary simulation with different fixed population sizes. Figure~\ref{fig:fig2} demonstrates that the final evolved strategy depends strongly on the population size. These results highlight that agents in smaller populations prefer risk averse strategies that receive a lower payoff but with higher reliability. In contrast, agents in larger populations do not show a preference for risk averse nor risk prone strategies and converge on $\bar\chi=0.5$ because all strategies perform roughly the same in large populations, that is, the $\chi$ of individual strategies drifts neutrally.

\subsection*{Theory of selection for variance}
The tendency for natural selection to select against variance in offspring number has been discussed before. Indeed, Gillespie has argued that large variance in offspring number could be selected against because adverse outcomes (few or zero offspring) cannot easily be balanced by favorable outcomes (large broods) for individuals of the same species, because the individuals without offspring may not get to try the ``offspring lottery" again~\citep{Gillespie1974,Gillespie1977}. Further, Gillespie proposed an approximate mean {\em actual} fitness that takes the variance in the offspring number distribution into account (see Methods, Equation~[\ref{correctFitness}]). 

\begin{figure}
\centering
\includegraphics[width=0.85\textwidth]{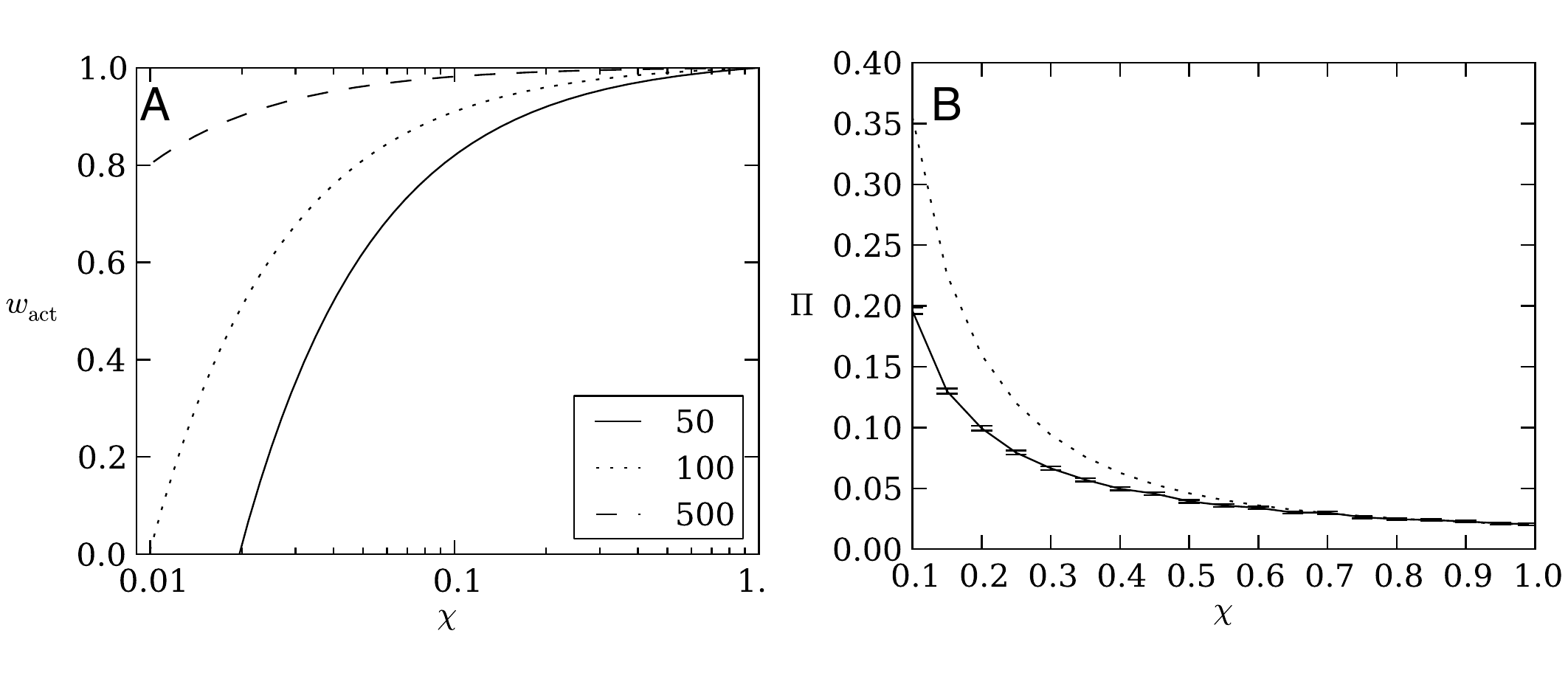}
\caption{A) Fitness ($w_{\rm act}$) as a function of strategy choice $\chi$ and population size according to Gillespie's model~Equation~(\ref{correctFitness}). Fitness differences between strategies with different $\chi$ are far more pronounced in smaller populations. Larger population sizes effectively buffer risky strategies against immediate extinction when the risky strategy does not pay off. See legend for population sizes. We note that the $x$-axis is log scale. B) Fixation probability ($\Pi$) of a perfectly risk-averse strategy ($\chi=1$) within a uniform background of strategies with choice $\chi$, as a function of $\chi$ (solid line). Fixation probabilities were estimated from 100,000 repeated runs, each seeded with a single invading strategy with $\chi=1$ in a background population of $N=50$ resident strategies with strategy $\chi$. The dashed line is Kimura's fixation probability $\Pi(s)=(1-e^{-2s})/(1-e^{-2Ns})$ (see, e.g.,~\citep{Gillespie2004}), where $s$ is the fitness advantage of the invading strategy calculated using Equation~(\ref{correctFitness}). Error bars are two standard errors.  
\label{fig:fig3}}
\end{figure}

Gillespie's fitness estimate $w_{\rm act}$ strongly depends on the strategy choice $\chi$ (because it determines the variance in the offspring number distribution) as well as population size (Figure~\ref{fig:fig3}A). Thus, his theory explains why agents that evolved in small populations show a preference for risk averse strategies, whereas agents that evolved in larger populations showed no such strategy preference. We can test the theory directly by measuring the probability $\Pi$ (the fixation probability) that a perfectly risk-averse strategy ($\chi=1$) can invade (and replace) a homogenous population consisting of strategies with choice $\chi\leq1$. We find 
that the observed probability of fixation (shown in Figure~\ref{fig:fig3}B for a population size of $N=50$) agrees qualitatively with the fixation probability calculated using Gillespie's fitness in Kimura's formula (dashed line in Figure~\ref{fig:fig3}B), but not quantitatively. Indeed, an effective fitness of about half of Gillespie's estimate reproduces the simulations almost exactly, which corroborates earlier findings~\citep{Shpak2005}.

\subsection*{Evolution in groups}

In the previous section, we demonstrated that agents in small populations evolve a preference for strategies with low variance in their payoff distribution, i.e., risk aversion. The group size for humans throughout evolutionary history has been proposed to be around 150 individuals~\citep{Aiello1993,Dunbar1993}, which suggests that evolving in such small groups could have been the reason behind the evolution of human risk aversion. However, a small group size and a small population size are two different things. While humans might have lived in small groups of 150 individuals, the total population size of humans has been much larger, and were only at times as low as 1,000 individuals~\citep{Cann1987,Vigilant1991}. Even though selection may occur within groups of about 150, individuals likely migrated between groups. Migration could have caused selection to effectively act on much larger groups (or even the entire human population) negating the selection for variance effect.

\begin{figure}[t]
\centering
\includegraphics[width=0.45\textwidth]{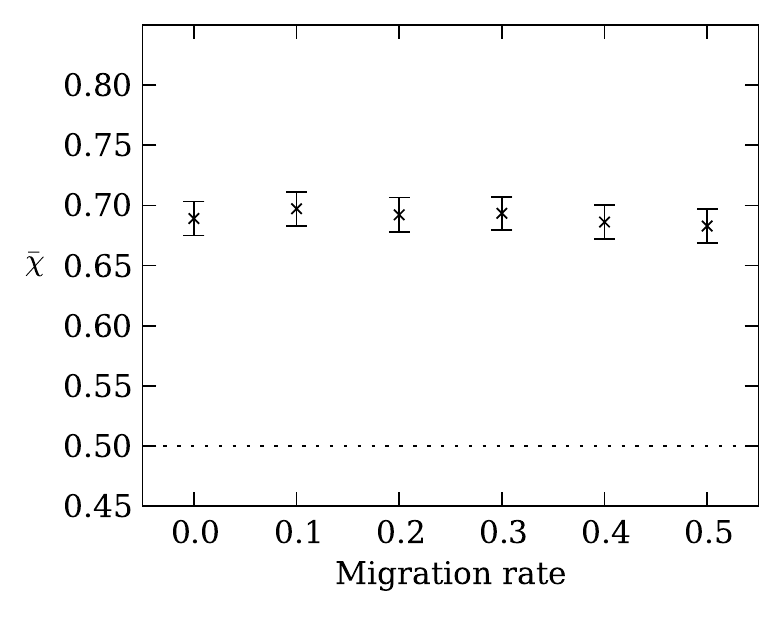}
\caption{Mean strategy $\bar \chi$ on the line of descent at generation 950 as a function of the migration rate in an island model genetic algorithm with 128 groups with 128 members in each group. Regardless of the migration rate, it is the group size and not the total population size that determines if the agents evolve risk averse strategies. Error bars indicate two standard errors over 1,000 replicates. The dotted line indicates the expected value for unbiased evolution, i.e., no strategy preference. A migration rate of 0.5 implies that half of the agents in each group migrate every generation. \label{fig:fig4}}
\end{figure}

We can simulate such an environment using an island-based evolutionary model (see Methods), in which individuals live in groups (the ``islands") that randomly exchange individuals with each other via migration. For example, we can run 1,000 replicate evolutionary experiments with 128 groups of 128 individuals each, with varying migration rates. In this configuration the total population size is 16,384 individuals, which according to Figure~\ref{fig:fig2} should result in agents evolving no strategy preference. Figure~\ref{fig:fig4} shows that regardless of the migration rate, the group size and not the total population size determines whether agents evolve risk averse strategies. This result suggests that even with migration between groups, the effective population size that selection acts on is determined by the group size and not the total population size.

\begin{figure}
\centering
\includegraphics[width=0.45\textwidth]{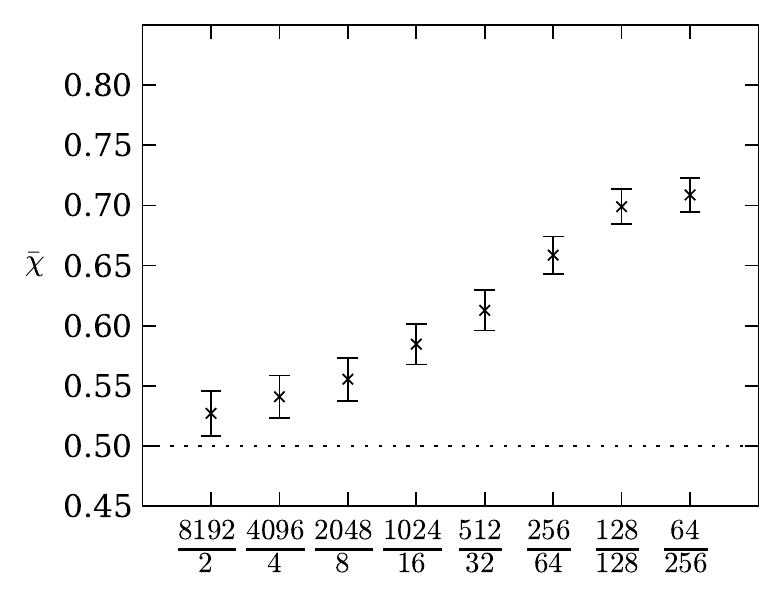}
\caption{Mean strategy $\bar \chi$ on the line of descent at generation 950 as a function of the ratio between group size and number of groups. Group size critically determines the preferred evolved agent strategies, where risk averse strategies are preferred in smaller groups and no strategy is preferred in larger groups. The $x$-axis tick labels are formatted as $\frac{{\rm group\ size}}{{\rm number\ of\ groups}}$. Error bars indicate two standard errors over 1,000 replicates. The dotted line indicates the expected value for unbiased evolution, i.e., no strategy preference.
\label{fig:sizeRatioIslandModel}} 
\end{figure}

When we change the size of the groups but fix the total population size (i.e., increase the group size and reduce the number of groups) while keeping the migration rate at a constant $0.1$, we again observe that the group size critically determines the preferred evolved agent strategies (Figure~\ref{fig:sizeRatioIslandModel}). Risk averse strategies are preferred in smaller groups and no strategy is preferred in larger groups. This result recapitulates the results from Figure~\ref{fig:fig2}, which shows that the preference for strategies with low payoff variation (i.e., risk aversion) depends on the effective population size.

\subsection*{Relative value of the gamble}

Another way to alter risk aversion in humans is by changing the relative value of the payoff~\citep{Holt2002}. When the gamble is about small amounts of money (i.e., ``peanuts" gambles or hypothetical money), humans tend to be less risk averse, whereas raising the relative value of the gamble increases risk aversion. In our evolutionary simulation, agents play the gamble a single time and the payoff they receive is their only source of fitness. This constraint effectively turns the gamble into a life or death situation, similar to a game with extraordinarily high stakes.

\begin{figure}[t]
\centering
\includegraphics[width=0.45\textwidth]{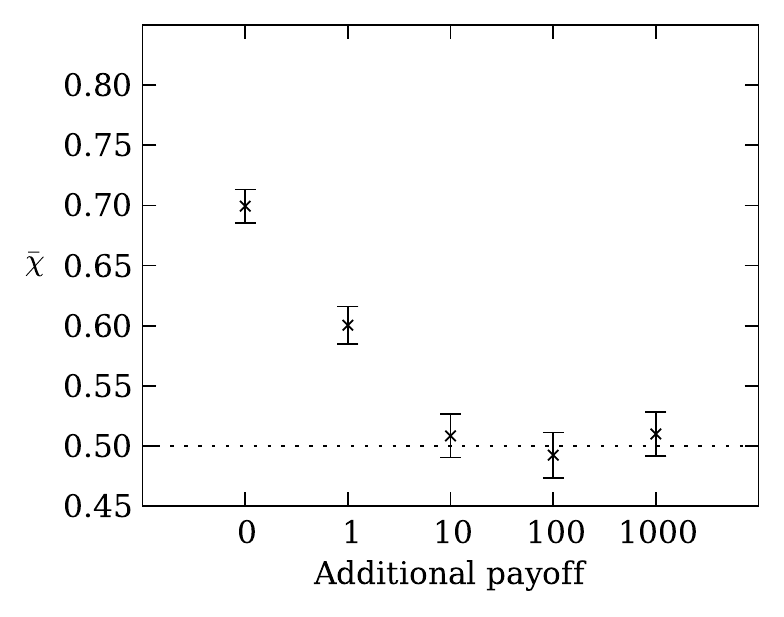}
\caption{Mean strategy $\bar \chi$ on the line of descent at generation 950 depending on the additional payoff ($\beta$). The larger the additional payoff $\beta$ becomes, the more often strategies return to an unbiased choice. Error bars indicate two standard errors over 1,000 replicates. The dotted line indicates the expected value for unbiased evolution, i.e., no strategy preference.
\label{fig:addonFitness}}
\end{figure}

To simulate lower-stakes gambles, we add a baseline payoff ($\beta$) to the payoff so that the fitness of the agent becomes
\be
X=\left\{    \begin{matrix} 
      \beta+\frac{1}{\chi} & p= \chi \\
      \beta & p=1-\chi\\
   \end{matrix}\right. \label{var2}\;
\ee
where $p$ is the probability to receive the corresponding payoff and $\chi$ is the agent's strategy. Typical gambles humans partake in fall either in the loss or in the gain domain. In biological systems, on the other hand, organisms accumulate resources in order to ultimately produce offspring. The ``gambles" these organisms undertake will influence the number of offspring, which will be positive or zero. Thus, we can not differentiate between losses or gains like humans would think about gambles for money. Therefore, gains and losses must be considered relative to fitness.

When we run the evolutionary simulation with a population size $N=100$ for various values of $\beta$, we observe that the larger the baseline $\beta$ becomes, the more often strategies return to an unbiased choice (Figure~\ref{fig:addonFitness}). This result is expected because fitness differences only matter if their relative impact is larger than $\frac{1}{N}$~\citep{Kimura1962,Gillespie2004}. Thus, risk averse strategies will only be selected for when the outcome of the gamble represents a significant portion of the individual's fitness when taking the population size into account.

\begin{figure}
\centering
\includegraphics[width=0.45\textwidth]{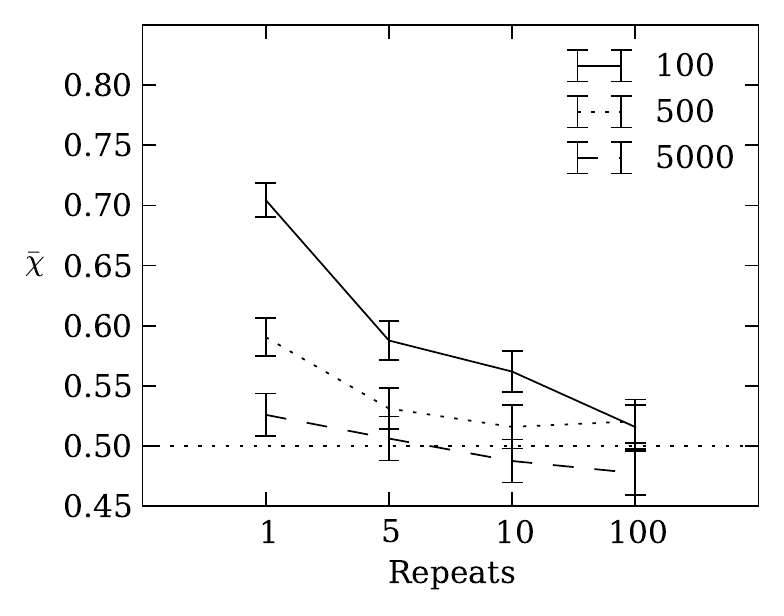}
\caption{Mean strategy $\bar \chi$  on the line of descent at generation 950 for three different population sizes (50, 500, and 5,000) depending on how many times the gambles are repeated. The more often the gamble is repeated during an individual's lifetime, the less likely risk aversion will evolve as a preferred strategy. Error bars indicate two standard errors over 1,000 replicates. The dotted line indicates the expected value $0.5$ for unbiased evolution, i.e., no strategy preference.
\label{fig:repeatedGames}}
\end{figure}

\subsection*{Repetition of the gamble}

Thus far, we have only investigated one-time gambles. What happens when the agents engage in the same gamble multiple times during their lifetime?  Intuitively, repeating the gamble should reduce the variance in the overall payoff the agents receive, and if gambles are played infinitely, the payoffs will converge on the same mean. In this experiment, we do not consider situations where agents can change their behavior based on previous experiences~\citep{Lopes1981}, but rather focus on unconditional responses. Shown in Figure~\ref{fig:repeatedGames}, we observe that the agents do not evolve a preference for risk aversion if the gamble is repeated several times in a lifetime. At the same time, this repetition effect depends strongly on the population size, such that smaller populations still evolve a preference for risk averse strategies with as many as 10 repetitions of the gamble. Therefore, a preference for risk aversion will only evolve for gambles that are encountered only a few times during an individual's lifetime. 

\section*{Discussion}

We hypothesized that risk aversion in humans could have been an evolutionary adaptation to living in small groups. We tested this hypothesis by evolving digital agents whose fitness is determined by a single choice during their lifetime in groups of varying size, and where that choice is encoded genetically and thus heritable. We observed that a preference for risk averse strategies does indeed evolve, but only when the group size is sufficiently small. These findings align with reports from earlier work that humans lived in groups of about 150 individuals for a large portion of their evolutionary history~\citep{Aiello1993,Dunbar1993}, providing a plausible evolutionary explanation for the risk averse behavior commonly observed in humans. In other words, these findings provide a quantitative foundation to the idea that evolution can explain risk aversion~\citep{Okasha2007}.

Additionally, we find that risk aversion is the preferred evolutionary adaptation to life in small groups when these groups are embedded within much larger groups, even with a large amount of migration between groups. However, it is important that the risky decisions occur only rarely during an individual's lifetime, and where the outcome of the risk represents a significant effect on the individual's fitness.  If the gamble has a negligible impact on fitness (e.g., only small gains are at stake) or if the risk is encountered regularly in the individual's lifetime, then the selective advantage of risk aversion will be lost. Examples of such rare, high-risk, high-payoff gambles include mating and mate competition~\citep{Buss1993}. 

Our work does not imply that no risk-prone strategies can possibly evolve. What we show is that risk aversion evolves on average, but the distribution of strategies within a population is quite broad (Figure~\ref{fig:figSingleLODpop1000beta1}). Thus, while on average agents are risk-averse if they evolve in a small population, there will always be some agents that are extremely risk-prone. Such agents can do extraordinarily well by chance and persist, but their genes are ultimately doomed for extinction.
 
While our model is only haploid and uses a single locus, we do not expect a diploid model using multiple loci to have qualitatively different results from the results presented in this paper. Regardless, gene flow in diploid organisms in an island model and its impact on the evolution of risk aversion is likely an interesting extension of this experiment to pursue in future work. 

When exploring whether risk aversion in human decisions today can be explained by evolution in the genes contributing to (risky) mate choice and mate competition during the species' evolutionary history, we must assume that genes involved in decisions that have a huge impact on fitness like mate choice are also involved in general decision making. This is supported by the evidence that the genes that subserve evaluation and reward are ancient~\citep{Symmondsetal2010}, suggesting that ancient risk averse behavior can still influence general decision mechanisms today. However, it is also clear that nature cannot be the only force that has shaped our risk averse behavior, because there is ample evidence that experience is a contributing factor. Thus, while this work has studied the impact that evolutionary history can have on strategy choice, it should be seen only as an element in understanding why humans shy away from risk.

\section*{Methods}
\subsection*{Single population evolutionary model}

We use a genetic algorithm applied on a population of agents to simulate evolution of the population~\citep{Michalewics1996}. Each agent in this population is defined by a probability $\chi$ (the ``choice"), which encodes the agent's strategy. We seed the initial agent population by assigning every agent a random $\chi$ drawn from a uniform distribution $(0, 1]$ with a variance of $\frac{1}{12}$. Varying the initial starting condition has no significant effect on the outcome of the experiments. Every agent in the population only plays the gamble once in its lifetime to determine its fitness, where $\chi$ is the probability to receive a fitness of $\frac{1}{\chi}$ or receive $0.0$ fitness with a chance $1-\chi$. The strategy of each agent can only change due to evolution, i.e., strategies cannot change during the agent's lifetime.

Once all of the agent fitnesses are evaluated for a given generation, the agents produce offspring into the next generation proportional to their fitness, i.e., we use fitness proportional roulette wheel selection to determine the next generation of individuals~\cite{Back1996}, implementing the Wright-Fisher process. Offspring inherit the strategy $\chi$ from their parent (no sexual recombination), except that 1\% of all offspring are subjected to mutation. If an offspring is subject to mutation, its new strategy is drawn randomly from a uniform distribution  $(0, 1]$. We repeat this evolutionary process every generation with a fixed population size for 1,000 generations.

\subsection*{Theory of selection for variance in offspring number}

Gillespie suggested that in finite populations where the fitness of individuals carries a stochastic component (modeled by a mean $\mu$ and a variance $\sigma^2$), the actual realized fitness $w_{\rm act}$ is given by~\citep{Gillespie1974,Gillespie1977}: 
\be
w_{\rm act}= \mu-\frac{1}{N}\sigma^2 \label{correctedFitness}\;,
\ee  
where $N$ is the population size. Because in the equivalent mean payoff gamble agents receive a payoff 
\be
X=\left\{    \begin{matrix} 
      \frac{1}{\chi} & p= \chi \\
      0 & p=1-\chi\\
   \end{matrix}\right. \label{var}
\ee   
the variance becomes 
\be
\sigma^2(X)=\frac{1}{\chi}-1 \label{variance}\;
\ee  
and the actual fitness of a strategy $\chi$ is
\be
w_{\rm act}(\chi)=1 -\frac{1}{N}(\frac{1}{\chi}-1) \label{correctFitness}\;,
\ee
as the mean of $X$ in Equation~(\ref{var}) (in an infinite population) equals 1. The fitness advantage $s$ of a strategy with $\chi=1$ versus a strategy $\chi$ is then
\be
s=\frac{w_{\rm act}(1)-w_{\rm act}(\chi)}{w_{\rm act}(\chi)}=\frac{1/\chi-1}{N-(1/\chi-1)}\;.  \label{adv}
\ee
We use Equation~(\ref{correctFitness}) to compute the actual fitness of a strategy using a given $\chi$ while taking the size of the population $N$ into account (Figure~\ref{fig:fig3}A), and we use the fitness advantage (\ref{adv}) in the calculation of the fixation probability using Kimura's formula in Figure~\ref{fig:fig3}B.

\subsection*{Island-based evolutionary model}

In our second set of experiments, we use an island-based evolutionary model to simulate an environment in which thousands of individuals are evolving in several small groups. For an overview of island models and the effect of population size, see~\citep{Whitley1998,Paz2003}. Island models have three parameters: The size of a single group, the number of groups, and a migration rate defining how many individuals per group are moved randomly to new groups during each generation. If an agent is selected to migrate, we randomly select a new group and a random agent within that group, and switch agents. Thus, our island-based evolutionary model implements several single population evolutionary models with a fixed fraction of individuals migrating between the populations every generation. The migration rate is the probability that an agent will be picked for migration per generation. For example, a migration rate of 0.1 implies that 10\% of the agents in the entire population are picked to switch (affecting up to 20\% of the population, as each switch affects two agents).

Typically, island models are used to speed up evolution in rugged fitness landscapes and increase genetic diversity within the population. In this experiment, we are not concerned with ruggedness nor diversity. Instead, we use an island model because it best approximates the scenario of individuals evolving in multiple small groups with some level of inter-group migration.

\subsection*{Retracing the line of descent}

At the end of each evolutionary run, we reconstruct the {\it line of descent} (LOD) by picking a random agent in the population and tracing back to the first generation using only direct ancestors~\citep{Lenski2003}. This procedure rapidly converges on the last most recent common ancestor (LMRCA) that swept the population. In our experiments, we determined that the agents on the LOD at generation 950 most often represented the LMRCA, thus we used those agents as the final representative agent for their respective evolutionary run. The LOD between the first agent and the LMRCA of a population contains all mutations that fixed during evolution, while all other mutants were outcompeted. Thus, analyzing an evolutionary run's LOD enables us to retrace the evolutionary history of the population.

\section*{Acknowledgments}
We thank Georg N\"oldeke for discussions and insights on risk aversion in economics, and Robert Heckendorn for discussions of island models. We gratefully acknowledge the support of the Michigan State University High Performance Computing Center and the Institute for Cyber Enabled Research (iCER).


\end{document}